\documentclass[12pt]{article}
\pdfoutput=1
\usepackage{amssymb,amsmath}
\usepackage{color,graphicx}
\usepackage{cite}
\usepackage{yfonts}
\newcommand\fverb{\setbox\pippobox=\hbox\bgroup\verb}
\newcommand\fverbdo{\egroup\medskip\noindent%
                              \fbox{\unhbox\pippobox}\ }
\newcommand\fverbit{\egroup\item[\fbox{\unhbox\pippobox}]}
\newbox\pippobox

\newcommand{\be} {\begin{equation}}
\newcommand{\ee} {\end{equation}}
\newcommand{\beq} {\begin{equation}}
\newcommand{\eeq} {\end{equation}}
\newcommand{\bea} {\begin{eqnarray}}
\newcommand{\eea} {\end{eqnarray}}
\newcommand{\bear}{\begin{eqnarray}}
\newcommand{\eear}{\end{eqnarray}}

\newcommand{\rc}{\nonumber\\}

\begin{document}
 
\begin{flushright}
HIP-2015-14/TH
\end{flushright}

\begin{center}

\centerline{\Large {\bf Cold holographic matter in the Higgs branch}}

\vspace{8mm}

\renewcommand\thefootnote{\mbox{$\fnsymbol{footnote}$}}
Georgios Itsios${}^{1,2}$\footnote{georgios.itsios@usc.es},
Niko Jokela${}^{3,4}$\footnote{niko.jokela@helsinki.fi},
and Alfonso V. Ramallo${}^{1,2}$\footnote{alfonso@fpaxp1.usc.es}

\vspace{4mm}
\vskip 0.2cm
${}^1${\small \sl Departamento de  F\'\i sica de Part\'\i  culas} \\
{\small \sl Universidade de Santiago de Compostela} \\
{\small \sl and} \\
${}^2${\small \sl Instituto Galego de F\'\i sica de Altas Enerx\'\i as (IGFAE)} \\
{\small \sl E-15782 Santiago de Compostela, Spain} 

${}^3${\small \sl Department of Physics} and ${}^4${\small \sl Helsinki Institute of Physics} \\
{\small \sl P.O.Box 64} \\
{\small \sl FIN-00014 University of Helsinki, Finland}

\end{center}

\vspace{8mm}
\numberwithin{equation}{section}
\setcounter{footnote}{0}
\renewcommand\thefootnote{\mbox{\arabic{footnote}}}

\begin{abstract}
\noindent
We study collective excitations of cold (2+1)-dimensional fundamental matter living on a defect of the four-dimensional ${\cal N}=4$ super Yang-Mills theory in the Higgs branch.  This system is realized holographically as a D3-D5 brane intersection, in which the D5-brane is treated as a probe with a non-zero gauge flux across the internal part of its worldvolume.  We study the holographic zero sound mode in the collisionless regime  at low temperature and find a simple analytic result for its dispersion relation. We also find the diffusion constant of the system in the hydrodynamic regime at higher temperature. In both cases we study the dependence on the flux parameter which determines the amount of Higgs symmetry breaking. We also discuss the anyonization of this construction.

\end{abstract}

\newpage

%
\renewcommand{\theequation}{{\rm\thesection.\arabic{equation}}}
%
%


\section{Introduction}
The gauge/gravity holographic duality has been recently employed to study compressible states of cold matter \cite{Karch:2008fa,Kulaxizi:2008kv,Davison:2011ek,Brattan:2012nb}.  In these studies the intersection of two different types of branes (D$p$ and D$q$ with $q\ge p$) is considered. The higher dimensional D$q$-branes are treated as probes  in the gravitational background generated by the lower dimensional D$p$-branes. 
On the field theory side the probe branes add hypermultiplets in the fundamental representation of the gauge group \cite{Karch:2002sh}.  These matter fields live generically in a defect of the unflavored theory. The probe approximation corresponds to the quenched approximation on the field theory side (see \cite{Jokela:2015aha} for an unifying formalism of the different brane intersections and for a complete list of references). 

In the brane setup the non-zero charge density needed to have a compressible state is generated  by turning on a worldvolume gauge field \cite{Kobayashi:2006sb}. The dominant collective excitation of these systems at sufficiently low temperatures is a sound mode (the holographic zero sound \cite{Karch:2008fa}). At  enough high temperature thermal effects dominate over quantum effects and the system enters in a hydrodynamic regime in which a diffusion mode dominates. These two regimes are connected by a collisionless/hydrodynamic crossover transition.

In this paper we will consider the 
intersection of D3- and D5-branes, according to the array:
\beq
\begin{array}{cccccccccccl}
 &0&1&2&3& 4& 5&6 &7&8&9 & \nonumber \\
D3: &\times & \times &\times &\times &\_ &\_ & \_&\_ &\_ &\_ &     \nonumber \\
D5: &\times &\times&\times&\_&\times&\times&\times&\_&\_&\_ &
\end{array}
\label{D3D5intersection}
\eeq
This D3-D5 system is
dual \cite{KR} to a defect theory in which ${\cal N}=4$, $d=4$ super
Yang-Mills theory in the bulk is coupled to ${\cal N}=4$, $d=3$
fundamental hypermultiplets localized at the defect \cite{WFO, EGK}.  We will restrict ourselves to the configuration in which the D3- and D5-branes are not separated in 789 directions, which corresponds to having massless hypermultiplet fields.  

Turning on a flux of the worldvolume gauge field along the internal directions 456, one realizes the Higgs branch of the theory \cite{Arean}, in which some components of the fundamental hypermultiplets acquire a non-vanishing vacuum expectation value. The worldvolume flux induces a bending of the D5-brane along the 3 direction. As shown in \cite{Arean} one can then regard the probe D5-brane as a bound state of D3-branes or, equivalently, one can interpret that  some of the D3-branes end on a D5-brane and recombine with it. The (2+1)-dimensional defect induced by the D5-brane represents a domain wall separating two regions with gauge groups with different ranks (the jump in the rank as one crosses the wall is proportional to the worldvolume flux).

In this paper we study the collective behavior of cold matter confined to the defect, when the system is in the Higgs branch. The first step  in our analysis will be determining the precise configuration of the probe which represents the system at non-zero chemical potential, temperature $T$, and magnetic field $B$ (the $T=B=0$ case was studied in \cite{Ammon:2012mu}).  We will then study the spectrum of excitations and we will determine the dispersion relation of the zero sound mode at $T=0$. We will find a simple analytical expression for the speed of zero sound as a function of the flux. Moreover, since the intersection is (2+1)-dimensional, we can consider mixed Dirichlet-Neumann boundary conditions in the UV, which corresponds to performing an alternative quantization \cite{Witten:2003ya} and thus the charge carriers become anyons. In the presence of the magnetic field the spectrum of the zero sound mode is generically gapped although, as in \cite{Jokela:2015aha}, one can adjust the anyon parameter to some critical value such that the resulting spectrum is gapless. We will also study the system at non-zero temperature and we will find the corresponding diffusion constant.

The rest of this paper is organized as follows. In section \ref{setup} we will determine precisely our brane configuration. The fluctuations of the D5-brane probe will be analyzed in section \ref{Fluctuations}. In section \ref{zero_sound} we obtain the spectrum of the zero sound. Section \ref{diffusion_constant} is devoted to the calculation of the diffusion constant. Finally, in section \ref{conclusions} we summarize our results and discuss some extensions of our work.

\section{The brane setup}
\label{setup}
Let us consider the supergravity solution corresponding to a stack of $N$ D3-branes at non-zero temperature. The corresponding near-horizon geometry is a black hole in $AdS_5\times {\mathbb S}^5$, whose metric is:
\beq
ds^2_{10}={r^2\over R^2}\Big(-f dt^2+d\vec x^2\Big)+{R^2\over r^2}\Big({dr^2\over f}+r^2 d\Omega_5^2\Big) \ ,
\label{10d_metric}
\eeq
where $\vec x=(x,y,z)$,  $R^4=4\pi g_s N\alpha'^2$ is the $AdS_5$ radius and the blackening factor
\beq
f(r)\,=\,1\,-\,{r_h^4\over r^4}\,\,.
\label{blackening}
\eeq
In (\ref{blackening}) $r_h$ is the horizon radius, related to the black hole temperature $T$ as
$r_h=\pi\, T$. 
The D3-brane background is endowed with a Ramond-Ramond five-form $F^{(5)}$, whose potential will be denoted by
$C^{(4)}$. The component of $C^{(4)}$ along the Minkowski coordinates
is given by:
\beq
\Big[\,C^{(4)}\,\Big]_{t, \vec x}\,=\,
{r^4\over R^4}\,\,.
\label{CRR}
\eeq
Let us now embed a D5-brane probe in the geometry (\ref{10d_metric}) in such a way that it is extended along $(x,y,r)$ and wraps a maximal ${\mathbb S}^2\subset {\mathbb S}^5$ (parameterized by two angles $\theta$ and $\varphi$).  If the D5-brane is bent along the third Cartesian coordinate $z=z(r)$, the induced metric on the worldvolume of the D5-brane is:
\beq
ds_6^2\,=\,r^2[\,-f\,dt^2\,+\,dx^2+dy^2]\,+\,\Big[{1\over r^2\, f}+r^2\,
z'^2\,\Big]\,dr^2\,+\, d\theta^2+\sin^2 \theta\,d\varphi^2\,\,,
\eeq
where we have taken units in which the $AdS_5$ radius $R=1$. We switch on a worldvolume gauge field given by:
\beq
F\,=\,A_t'\, dr \wedge dt  \,+\,B\, dx\wedge dy+\,q\sin\theta\,d\theta\wedge d\varphi\,\,,
\label{F_unperturbed}
\eeq
with $q$ being a constant (the amount of  flux).\footnote{The flux number $q$ must satisfy the following quantization condition, $q=\pi \alpha' k$, with $k\in {\mathbb Z}$ (see, for example, \cite{Arean}). However, in  units in which $R=1$, the Regge slope is $\alpha'=1\sqrt{4\pi N g_s}$. Accordingly, we will consider $q$ as a continuous parameter.} As usual, the $r\, t$ component of $F$ in (\ref{F_unperturbed}) is required in order to have a non-vanishing charge density.  The action of a D5-brane probe in the  background  geometry is
given by the sum of the Dirac-Born-Infeld (DBI) and Wess-Zumino (WZ) terms:
\beq
S_{D5}\,=\,-\,T_{5}\,\int d^6\xi\,\sqrt{-\det (g+F)}\,+\,
T_{5}\,\int d^6\xi \,\,\,\hat C^{(4)}\,\wedge F\,\,,
\label{DBI-D5}
\eeq
where $T_{5}$ is the tension of the D5-brane and $g$ is the induced  metric on the worldvolume. 
The DBI determinant for our ansatz  is:
\beq
\sqrt{-\det (g+F)}\,=\,\sqrt{r^4+B^2}\,\sqrt{1+r^4\,f\,z'^2\,-\,A_t'^{\,2}}\,
\sqrt{1+q^2}\,\sin\theta\,\,,
\eeq
while the WZ Lagrangian density is given by:
\beq
{\cal L}_{WZ}\,=\,T_5\,\hat C_4\,\wedge\,F\,=\,T_5\,q\,r^4\,z'\,\sin\theta\,
dt\wedge dx\wedge dy\wedge dr\wedge d\theta \wedge d\varphi\,\,.
\eeq
Therefore, after integrating over the angular variables, the total Lagrangian density becomes:
\beq
{\cal L}\,=\,4\pi T_5\,\Big[-\sqrt{r^4+B^2}\,\sqrt{1+r^4\,f\,z'^2\,-\,A_t'^2}\,
\sqrt{1+q^2}+q\,r^4\,z'\Big]\,\,,
\eeq
The equation of motion for $A_t$ leads to:
\beq
{\sqrt{r^4+B^2}\,A_t'\over 
\sqrt{1+r^4\,f\,z'^2\,-\,A_t'^2}}\,\sqrt{1+q^2}\,=\,d\,\,,
\eeq
where $d$ is a constant proportional to  the charge density. From this equation we get $A_t'$ as a function of $z'$:
\beq
A_t'\,=\,{d\,\sqrt{1+r^4\,f\,z'^2}\over 
\sqrt{d^2+(1+q^2)\,(r^4+B^2)}}\,\, .
\label{At_prime_z_prime}
\eeq
The equation of motion for $z$ leads to the equation:
\beq
-\sqrt{r^4+B^2}\,{r^4\,f\,z'\over 
\sqrt{1+r^4\,f\,z'^2\,-\,A_t'^{\,2}}}\,\sqrt{1+q^2}\,+\,q\,r^4\,=\,c_z\,\,,
\label{eom_z}
\eeq
where $c_z$ is a constant of integration. By imposing regularity at the horizon of the embedding function $z(r)$ \cite{Bergman:2010gm}, yields $c_z\,=\,q\,r_h^4$. It is then possible to use (\ref{At_prime_z_prime}) and (\ref{eom_z}) to obtain
$A_t'$ and $z'$ as functions of the coordinate $r$:
\beq
A_t'\,=\,{d\over \sqrt{r^{4}+d^2+q^2\,r_h^4+(1+q^2)\,B^2}}\,\,,
\qquad\qquad
z'\,=\,{q\over \sqrt{r^{4}+d^2+q^2\,r_h^4+(1+q^2)\,B^2}}\,\,.
\qquad
\label{At_z_prime_explicit}
\eeq

In what follows it is convenient to express the different results in terms of the chemical potential $\mu$ at $T=B=0$, which is given by:
\beq
\mu=A_t(r=\infty)=\int_0^{\infty}\,dr\,A_t'(r){\big|_{T=B=0}}={4\,\Gamma\big({5\over 4}\big)^2\over \sqrt{\pi}}\,d^{{1\over 2}}\,\,.
\label{chemical_pot}
\eeq

\section{Fluctuations}
\label{Fluctuations}

Let us  allow fluctuations of the gauge field along the Minkowski directions of the intersection, in the form:
\beq
A\,=\,A^{(0)}\,+\,a (r, x^{\mu})\,\,,
\eeq
where  $A^{(0)}$ is the gauge potential for the field strength (\ref{F_unperturbed}) and  $a(r,x^{\mu})=a_{\nu}(r,x^{\mu})\,dx^{\nu}$ is a fluctuation.  The total gauge field strength is:
\beq
F_{ab}\,=\,F^{(0)}_{ab}\,+\,f_{ab}\,\,,
\eeq
with  $F^{(0)}=dA^{(0)}$ is the two-form written in (\ref{F_unperturbed}). In order to write the Lagrangian for the fluctuations at second order,  let us split the inverse of the matrix $g^{(0)}\,+\,F^{(0)}$ as:
\beq
\Big(\,g^{(0)}\,+\,F^{(0)}\Big)^{-1}\,=\,{\cal G}^{-1}\,+\,{\cal J}\,\,,
\eeq
where ${\cal G}^{-1}$ is the symmetric part and ${\cal J}$ is the antisymmetric part (${\cal G}$ is the so-called open string metric). Then,  the Lagrangian density for the fluctuations is:
\beq
{\cal L}\,\sim\,{r^4+B^2\over \sqrt{r^{4}+d^2+q^2\,r_h^4+(1+q^2)\,B^2}}\,\,\,\,
\Big(\,{\cal G}^{ac}\,{\cal G}^{bd}\,-\,
{\cal J}^{ac}\,{\cal J}^{bd}\,+\,{1\over 2}\,{\cal J}^{cd}\,{\cal J}^{ab}
\Big)f_{cd}\,f_{ab}\,\,,
\label{Lag_fluct}
\eeq
where the Latin indices take values in $a,b,c\in \{t,x,y,r\}$. Notice that we are choosing a gauge in which $a_r=0$. The  equation of motion  for $a^d$ derived from (\ref{Lag_fluct})  is:
\beq
\partial_{c}\,\Bigg[{r^4+B^2\over
\sqrt{r^{4}+d^2+q^2\,r_h^4+(1+q^2)\,B^2}
}\,
\Big(\,{\cal G}^{ca}\,{\cal G}^{db}\,-\,
{\cal J}^{ca}\,{\cal J}^{db}\,+\,{1\over 2}\,{\cal J}^{cd}\,{\cal J}^{ab}
\Big)\,f_{ab}\Bigg]\,=\,0\,\,.
\label{eom_general}
\eeq
Let us write these equations in the case in which the fluctuation fields $a_{\nu}$ only depend on 
the coordinates $r$, $t$, and $x$. We first Fourier transform  the gauge field to momentum space as:
\beq
a_\nu(r, t, x)\,=\,\int {d\omega\,dk\over (2\pi)^2}\,
a_\nu(r, \omega, k)\,e^{-i\omega t+i k x}\,\,.
\eeq
In what follows it will be understood that the gauge field is written in momentum space. 
Moreover, we define the electric field $E$ as the gauge-invariant combination:
\beq
E\,=\,k\,a_t\,+\,\omega\,a_x\,\,.
\label{E_at_ax}
\eeq
The equations of motion reduce to a set of two coupled equations for $E$ and the transverse gauge field fluctuation $a_y$.  The equation for the fluctuation of the electric field $E$ is given by:
\bear
&&E''+\,\partial_r\log\Bigg[
{r^{4} f\over r^{4}+B^2}\,
{\sqrt{r^{4}+d^2+q^2\,r_h^4+(1+q^2)\,B^2}\,
\Big(d^2+(1+q^2)\,(r^{4}+B^2)\Big)
\over
(1+q^2)(\omega^2-\,f\, k^2)r^{4}+[(1+q^2)\,B^2\,+\,d^2]\,\omega^2}
\Bigg]\,E'\rc
&&\qquad\qquad
+{1\over r^4 f^2}\,{
(1+q^2)(\omega^2-\,f\, k^2)r^{4}+[(1+q^2)\,B^2\,+\,d^2]\,\omega^2
\over 
r^{4}+d^2+q^2\,r_h^4+(1+q^2)\,B^2
}\,E\rc
&&=-{4i Bd\over  r\,(r^{4}+B^2)f}\,
{(1+q^2)(\omega^2-\,f\, k^2)r^{4}+[(1+q^2)\,B^2\,+\,d^2]\,\omega^2
\over 
\sqrt{r^{4}+d^2+q^2\,r_h^4+(1+q^2)\,B^2}\,
\Big(d^2+(1+q^2)\,(r^{4}+B^2)\Big)
}\,a_y\,\,.
\label{eom_E}
\eear
The equation for the transverse fluctuation is:
\bear
&&a_y''\,+\,\partial_r\log\Bigg[{r^{4}\,f\over r^{4}+B^2}\,
\sqrt{r^{4}+d^2+q^2\,r_h^4+(1+q^2)\,B^2}
\Bigg]\,a_y'\rc
&&\qquad\qquad
+{1\over r^{4}f^2}\,{
(1+q^2)(\omega^2-\,f\, k^2)r^{4}+[(1+q^2)\,B^2\,+\,d^2]\,\omega^2
\over 
r^{4}+d^2+q^2\,r_h^4+(1+q^2)\,B^2}\,a_y\rc
&&\qquad\qquad
=\,{4i Bd\over  r\,(r^{4}+B^2)f}\,
{E\over 
\sqrt{r^{4}+d^2+q^2\,r_h^4+(1+q^2)\,B^2}}\,\,.
\label{eom_ay}
\eear
Let us now see how one can eliminate the dependence of $r_h$ on the equations of motion by performing appropriate rescalings. First of all, we define a new radial variable $\hat r\,=\,r/ r_h$. 
Then, one can check that $r_h$ is eliminated from (\ref{eom_E}) and (\ref{eom_ay})  by defining the new rescaled (hatted) quantities as:
\beq
\hat \omega\,=\,{\omega\over r_h}\,\,,
\qquad\qquad
\hat k\,=\,{k\over r_h}\,\,,
\qquad\qquad
\hat d\,=\,{d\over r_h^2}\,\,,
\qquad\qquad
\hat B\,=\,{B\over r_h^{2}} \ .
\label{rescaled_quantities}
\eeq
The resulting equations are obtained from (\ref{eom_E}) and (\ref{eom_ay})  by taking $r_h=1$ and substituting all quantities by their hatted counterparts. Hatted variables are utilized  in the numerical integration of eqs. (\ref{eom_E}) and (\ref{eom_ay}).

\section{Zero sound}
\label{zero_sound}

Let us now study the system at zero temperature. First we study the equations of motion (\ref{eom_E}) and (\ref{eom_ay}) near the Poincar\'e horizon $r=0$. Assuming that $B$ is small ($B\sim r^4$), the equations of $E$ and $a_y$ are given by the coupled system:
\bear
E''\,+\,{4\,B^2\over r(r^{4}+B^2)}\,E'\,+\,{\omega^2\over r^{4}}\,E & = &
-\,{4i\,B\omega^2\over r (r^{4}+B^2)}\,a_y \rc
a_y''\,+\,{4\,B^2\over r(r^{4}+B^2)}\,a_y'\,+\,{\omega^2\over r^{4}}\,a_y & = &
\,{4i\,B\over r (r^{4}+B^2)}\,E\,\,.
\label{E_ay_coupled}
\eear
This is the same system as in the D3-D5 case with zero flux of \cite{Brattan:2012nb}. We can readily write its solution in matrix form as:
\beq
\begin{pmatrix}
E\\ \\ a_y
\end{pmatrix}
\,=\,e^{{i\omega\over r}}\,
\begin{pmatrix}
  r &&& \big(1-{i\omega\over r}\big)\,{B\over \omega}\\
  {} & {} \\
  {i\over \omega}\,\big(1-{i\omega\over r}\big)\,{B\over \omega} &&& {i\over \omega}\,r
 \end{pmatrix}\,
 \begin{pmatrix}
c_1\\ \\c_2
\end{pmatrix}\,\,,
\label{nh_solution}
\eeq
where $c_1$ and $c_2$ are  integration constants and we have imposed infalling boundary conditions at the horizon. Next we take the limit of low frequency and momentum in such a way that $\omega\sim k\sim {\cal O}(\epsilon)$.  For small $\omega$ the solution (\ref{nh_solution}) can be written as:
\beq
\begin{pmatrix}
E\\ \\ a_y
\end{pmatrix}
\,=\,
\begin{pmatrix}
  r+i\omega &&& \,{B\over \omega}\,
  {} & {} \\\\
  {iB\over \omega^2} &&& {i\over \omega}\,(r+i\omega)
 \end{pmatrix}\,
 \begin{pmatrix}
c_1\\ \\c_2
\end{pmatrix}\,\,.
\label{D3_D5_nh_low_freq_B}
\eeq
We now take the low frequency limit first. One can verify that the equations decouple in this limit. Actually, the equation for $E$ becomes:
\beq
E''+\partial_r\log{
\sqrt{r^{4}+d^2}\,[d^2+(1+q^2)\,r^{4}]\over
(1+q^2)(\omega^2-\, k^2)r^{4}+\omega^2\,d^2}\,E'\,=\,0\,\,.
\eeq
This equation can be readily integrated as:
\beq
E(r)\,=\,E^{(0)}\,-\,c_E\,
\int_{r}^{\infty}\,d\rho\,
{(1+q^2)(\omega^2-\, k^2)\rho^{4}+\omega^2\,d^2\over
\sqrt{\rho^{4}+d^2}\,[d^2+(1+q^2)\,\rho^{4}]}\,\,,
\label{E_lowfreq}
\eeq
where $E^{(0)}=E(r\to\infty)$. Actually, 
if we define the integrals:
\bear
&& {\cal K}_1(r)\,=\,\int_{r}^{\infty}\,
{\rho^4\over \sqrt{\rho^{4}+d^2}\,[d^2+(1+q^2)\,\rho^{4}]}\,\,d\rho\rc
&& {\cal K}_2(r)\,=\,\int_{r}^{\infty}\,
{d\rho\over \sqrt{\rho^{4}+d^2}\,[d^2+(1+q^2)\,\rho^{4}]}\,\,,
\eear
then, $E(r)$ can be written as:
\beq
E(r)\,=\,E^{(0)}\,-\,c_E\,\Big[(1+q^2)(\omega^2-\, k^2)\,{\cal K}_1(r)\,+\,
\omega^2\,d^2\,{\cal K}_2(r)\Big]\,\,.
\eeq
Let us now expand  this result near the horizon. We first expand the integrals ${\cal K}_1(r)$ and
$ {\cal K}_2(r)$ near $r=0$ as:
\beq
 {\cal K}_1(r)\,=\,\bar  {\cal K}_1\,+\,{\mathcal O}(r^2)\,\,,
 \qquad\qquad
 {\cal K}_2(r)\,=\,\bar  {\cal K}_2\,-{r\over d^3}\,+\,{\mathcal O}(r^2)\,\,,
 \eeq
where $\bar  {\cal K}_1={\cal K}_1(r=0)$ and $\bar  {\cal K}_2={\cal K}_2(r=0)$. It is interesting to notice that the quantities  $\bar  {\cal K}_1$ and $\bar  {\cal K}_2$ are not independent. Indeed, they satisfy the relation:
\beq
(1+q^2)\bar  {\cal K}_1\,+\,d^2\,\bar  {\cal K}_2\,=\,\int_0^{\infty}
{d\rho\over \sqrt{\rho^4+d^2}}\,=\,
{\mu\over d}\,\,.
\eeq
Moreover, if we define the flux function ${\cal F}(q)$ as:
\beq\label{flux_function}
{\cal F}(q)\,\equiv\,{2d\over \mu}\,(1+q^2)\,\bar  {\cal K}_1=(1+q^2)\,F\Big(1,{5\over 4};{3\over 2};-q^2\Big) \ ,
\eeq
then, near $r=0$ we can write $E(r)$ as:
\beq
E\,=\,c_E\,{\omega^2\over d}\,r\,+\,E^{(0)}\,-\,{c_E\over d}\,
\Big[\mu\,\omega^2\,-\,{\mu\over 2}\,{\cal F}(q)\,k^2\Big]\,\,.
\label{E_lowfreq_nh}
\eeq
It is worth stressing that the whole effect of the flux in (\ref{E_lowfreq_nh}) is equivalent to multiplying $k^2$ by the flux function ${\cal F}(q)$.

The equation for $a_y$ for low frequency is:
\beq
a_y''\,+\,\partial_{r}\,\log\Big(r^{4}+d^2\Big)^{{1\over 2}}\,a_y'= 0\,\,.
\eeq
This equation can be integrated twice to give:
\beq
a_y(r)\,=\,a_y^{(0)}\,-\,c_y
\int_{r}^{\infty}\,
{d\rho\over
(\rho^{4}+d^2)^{{1\over 2}}}\,=\,
a_y^{(0)}\,-\,{c_y\over r}\,
F\Big(\,{1\over 2}, {1\over 4}; {5\over 4};
- {d^2\over r^{4}}\,\Big)\
\,\,,
\label{ay_low_frequency}
\eeq
with $a_y^{(0)}\,=\,a_y(r\to\infty)$. 
For small $r$ the previous solution becomes:
\beq
a_y(r)\,\approx\,a_y^{(0)}\,-\,c_y\,{\mu\over d}\,+\,c_y\,{r\over d}\,\,.
\label{ay_low_nh}
\eeq
Let us now match the two expressions we have found for $E$ and $a_y$ in this double  low frequency and near-horizon limit (eqs. (\ref{D3_D5_nh_low_freq_B}), (\ref{E_lowfreq_nh}), and (\ref{ay_low_nh})).  Looking at the terms linear in $r$ we get find relations between the constants $c_1$, $c_2$,  $c_E$, and $c_y$:
\beq
c_1\,=\,{\omega^2\over d}\,c_E\,\,,
\qquad\qquad\qquad
c_2\,=\,-i{\omega\over d}\,c_y\,\,.
\eeq
The identification of the constant terms yields the following matrix relation
\beq
\begin{pmatrix}
E^{(0)}\\ \\ a_y^{(0)}
\end{pmatrix}
\,=\,
\begin{pmatrix}
 i{\omega^3\over d}\,+\,{\mu\over d}\,\omega^2\,-\,
{\mu\over 2d}\,{\cal F}(q)\,k^2
 &&&-i {B\over d}\\
  {} & {} \\
  {iB\over d} &&& {i\omega\over d}\,+\,{\mu\over d}
 \end{pmatrix}\,
 \begin{pmatrix}
c_E\\ \\c_y
\end{pmatrix}\,\,.
\label{E0_ay0_B}
\eeq

\begin{figure}[ht]
\center
 \includegraphics[width=0.42\textwidth]{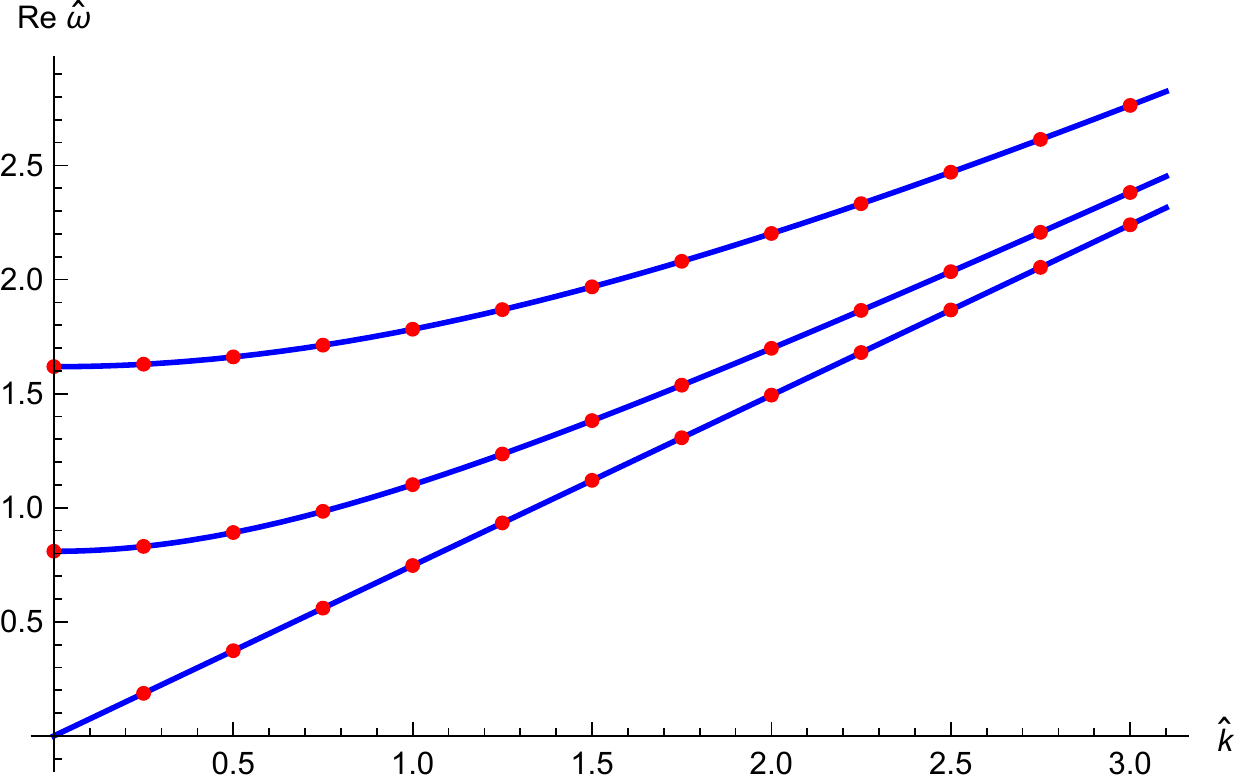}
 \qquad\qquad
 \includegraphics[width=0.42\textwidth]{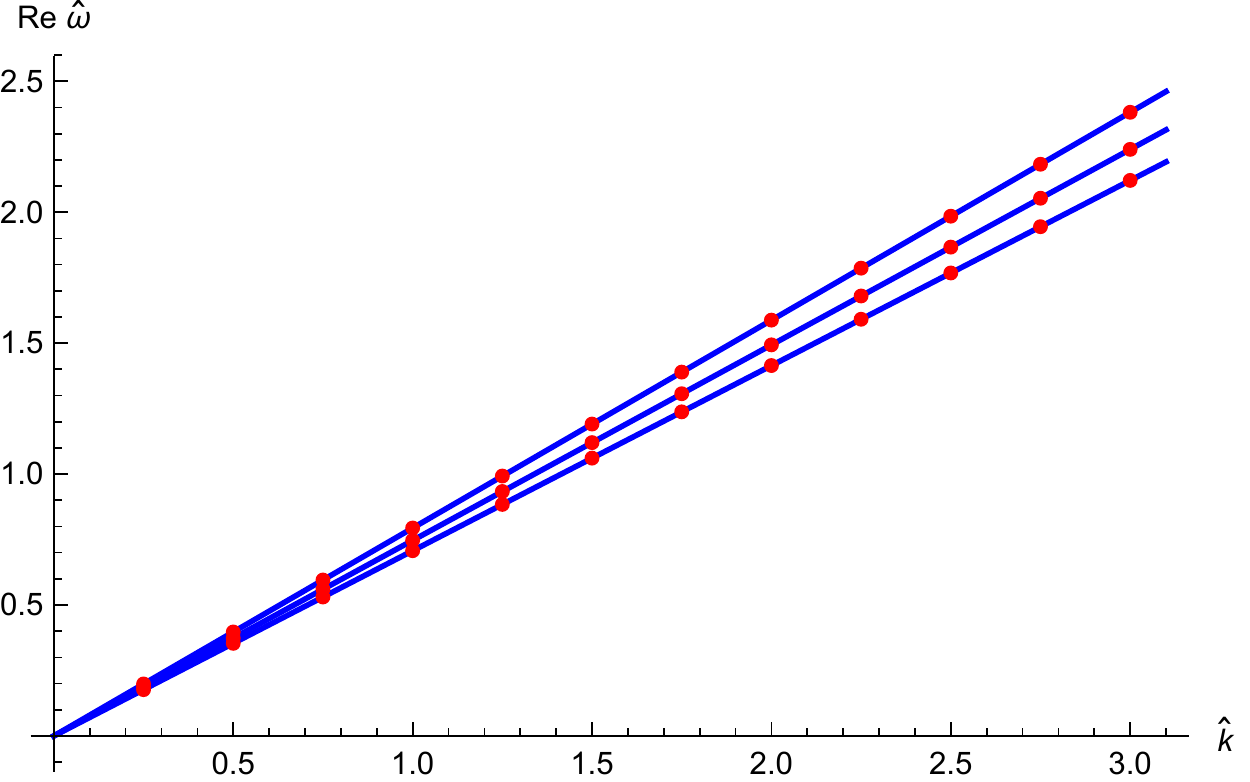}
 \caption{(left) The dispersions for the anyon fluid as the effective magnetic field is decreased to zero. The red points are the numerical values for $q=1$ and $n=0,n_{crit}/2,n_{crit}$ (top-down). (right) The dispersions for the anyon fluid at criticality $n=n_{crit}$ with increasing flux $q=0,1,2$ (bottom-up). Both figures are done at $\hat d=10^6$, $\hat B=3\cdot 10^3$.}
 \label{dispersion_k}
\end{figure}

\begin{figure}[ht]
\center
 \includegraphics[width=0.42\textwidth]{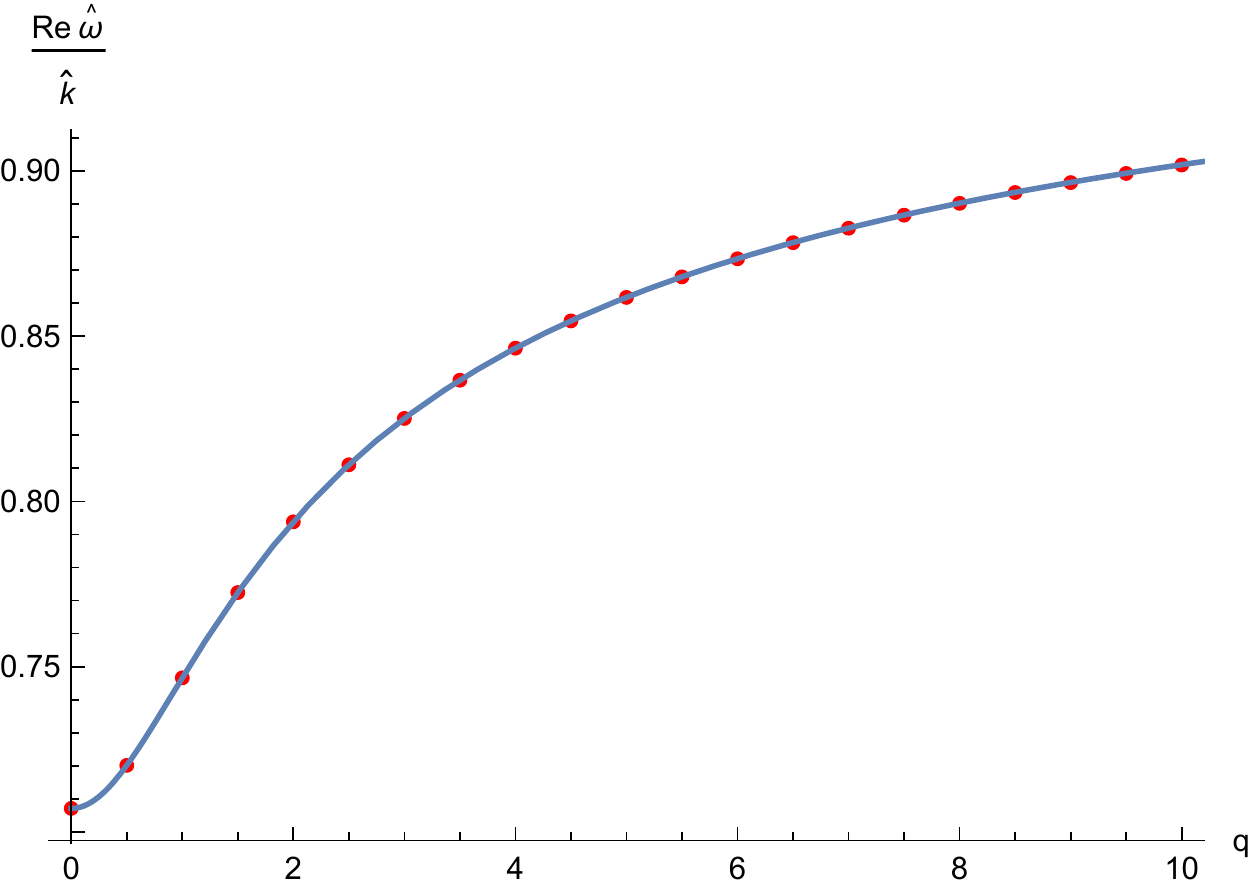}
 \qquad\qquad
 \includegraphics[width=0.42\textwidth]{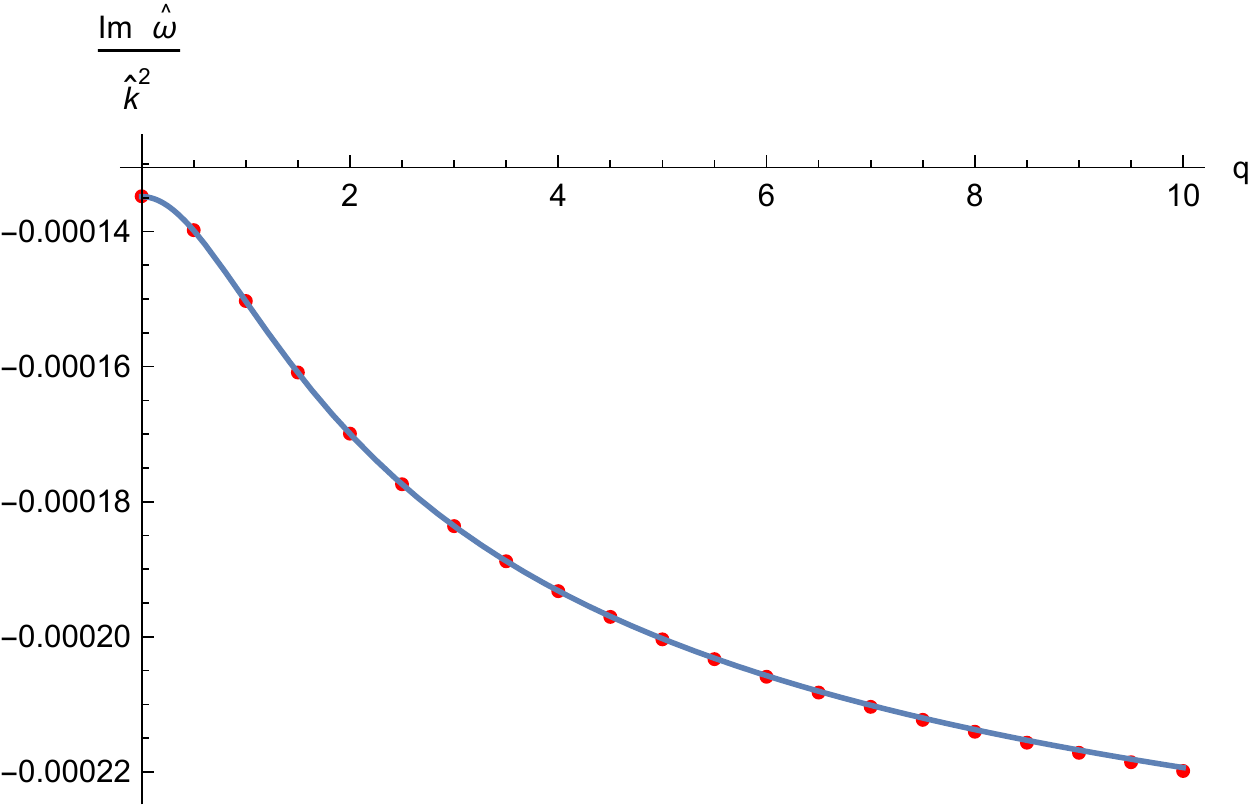}
 \caption{We display the dispersions of the lowest quasinormal mode at low temperature (zero sound) as functions of flux $q$. The analytic formulas (\ref{Re_omega}) and (\ref{Im_omega}) (blue curves) match extremely well with numerics (red points) for the real and imaginary parts of the frequency $\hat \omega$, respectively, as long as the temperature is kept small enough; we use $\hat d=10^6$ and $\hat k=10$ with $\hat B=\,n\,=0$. }
  \label{dispersion_q}
\end{figure}

Let us  now require that our fluctuation modes satisfy the following mixed Dirichlet-Neumann boundary conditions at the UV\cite{Jokela:2013hta,Brattan:2013wya}:
\beq
\lim_{r\to\infty}\,\Big[\,n\,r^{\,2}\,f_{r\,\mu}\,-\,{1\over 2}\,
\epsilon_{\mu\alpha\beta}\,f^{\alpha\beta}\,\big]\,=\,0\,\,,
\label{bc_with_n}
\eeq
with $n$ being some constant (the Dirichlet boundary conditions correspond to taking $n=0$). As in the case with $q=0$, these conditions are equivalent to:
\beq
\lim_{r\to\infty}\,E\,=\,-i\,n\,\lim_{r\to\infty}\,\big[\,r^{\,2}\,a_y'\,\big]\,\,,
\qquad\qquad\qquad
\lim_{r\to\infty}\,a_y\,=\,
i\,{n\over \omega^2-k^2}\,\lim_{r\to\infty}\,\big[\,r^{\,2}\,E'\,\big]\,\,.
\label{lim_E_ay}
\eeq
The quantities on the left-hand-side of (\ref{lim_E_ay}) are the UV values $E^{(0)}$ and $a_y^{(0)}$. In order to obtain the values of the right-hand-side of the two conditions in  (\ref{lim_E_ay}), notice that the radial derivatives of $E$ and $a_y$ can be obtained from their low-frequency values (\ref{E_lowfreq}) and (\ref{ay_low_frequency}):
\beq
{\partial E\over \partial r}\,\Big|_{r\to\infty}\approx (\omega^2-k^2)\,c_E\,r^{-2}\,\,,
\qquad\qquad
{\partial a_y\over \partial r}\,\Big|_{r\to\infty}\approx c_y\,
r^{-2}\,\,.
\eeq
From these expressions we can recast the boundary conditions (\ref{bc_with_n}) as relations between the constants $E^{(0)}$, $a_y^{(0)}$, $c_E$, and 
$c_y$. Indeed, let us define  $E_{n}^{(0)}$ and $a_{y,n}^{(0)}$ as:
\beq
E_{n}^{(0)}\,\equiv E^{(0)}\,+\,i\,n\, c_y\,\,,
\qquad\qquad
a_{y,n}^{(0)}\,=\,a_y^{(0)}\,-\,i\,n\,c_E\,\,.
\label{E_n-ay_n}
\eeq
Then, (\ref{lim_E_ay}) is equivalent to the conditions:
\beq
E_{n}^{(0)}\,=\,a_{y,n}^{(0)}\,=\,0\,\,.
\label{En0_ay_n0_def}
\eeq
Moreover, combining (\ref{E0_ay0_B}) and (\ref{E_n-ay_n}) we conclude that 
$E_{n}^{(0)}$ and $a_{y,n}^{(0)}$ are related to the constants $c_E$ and $c_y$ by the following matrix equation:
\beq
\begin{pmatrix}
E_{n}^{(0)}\\ \\ a_{y,n}^{(0)}
\end{pmatrix}
\,=\,
\begin{pmatrix}
 i{\omega^3\over d}\,+\,{\mu\over d}\,\omega^2\,-\,
{\mu\over 2d}\,{\cal F}(q)\,k^2
 &&&-i {B\over d}+in\\
  {} & {} \\
  {iB\over d}-in &&& {i\omega\over d}\,+\,{\mu\over d}
 \end{pmatrix}\,
 \begin{pmatrix}
c_E\\ \\c_y
\end{pmatrix}\,\,.
\label{En0_ayn0_B}
\eeq
Furthermore, the non-trivial fulfillment of the condition (\ref{En0_ay_n0_def}) is equivalent to
the vanishing of the determinant of the matrix in (\ref{En0_ayn0_B}), which  determines the dispersion relation satisfied by $\omega$ and $k$ for the zero sound modes:
\beq
\omega^4\,-\,2i\mu\,\omega^3\,+\,i\,{\mu\over 2}\,{\cal F}(q)\,\omega\,k^2\,-\,\mu^2\,\omega^2\,+\,
{\mu^2\over 2}\,{\cal F}(q)\,k^2\,+\,(B-nd)^2\,=\,0\,\,.
\label{disp_rel_B_flux}
\eeq
Notice that the effect of the flux in (\ref{disp_rel_B_flux}) is encoded in the substitution  
$k^2\to {\cal F}(q)\,k^2$. 
Let us now solve (\ref{disp_rel_B_flux}) for $\omega$ as a function of $k$ for small values of $(\omega, k)$.  At leading order $\omega$ is real and given by:
\beq
\omega^2\,=\,
{1+q^2\over 2}\,F\Big(1,{5\over 4};{3\over 2};-q^2\Big)\,
\,k^2\,+\,{(B-nd)^2\over \mu^2}\,\,.
\label{Re_omega}
\eeq
It follows from the last term in (\ref{Re_omega}) that the spectrum is generically gapped for non-vanishing  $B$ and $n$. However, it can be made gapless by adjusting the alternative quantization parameter $n$ to the critical value $n_{crit}=B/d$.  This fact is illustrated in Fig.~\ref{dispersion_k}, where we compare the numerical results to our analytic formula (\ref{Re_omega}).  Moreover, from the coefficient of the momentum in the right-hand side of 
(\ref{Re_omega}) we can extract the dependence of the speed of zero sound  $u_0$ on the flux $q$. Indeed, by analyzing the behavior of the flux function (\ref{flux_function}),  it is easy to conclude that  $u_0=\pm 1/\sqrt{2}$ when $q=0$, whereas it approaches the maximal possible value $u_0=\pm 1$ as $q\to\infty$. Moreover, solving (\ref{disp_rel_B_flux}) for $\omega$ at next-to-leading order, we find the attenuation of the zero sound:
\beq
{\rm Im}\,\omega\,=\,-{1\over \mu}\,\Big[
{1+q^2\over 4 }\,F\Big(1,{5\over 4};{3\over 2};-q^2\Big)\,
\,k^2\,+\,{(B-nd)^2\over \mu^2}\Big]\,\,.
\label{Im_omega}
\eeq
In Fig.~\ref{dispersion_q} we compare the analytic results ${\rm Re}\,\omega$ and ${\rm Im}\,\omega$ with the numerics with varying flux $q$ and find very good agreement.

\section{Diffusion constant}
\label{diffusion_constant}
\begin{figure}[ht]
\center
 \includegraphics[width=0.6\textwidth]{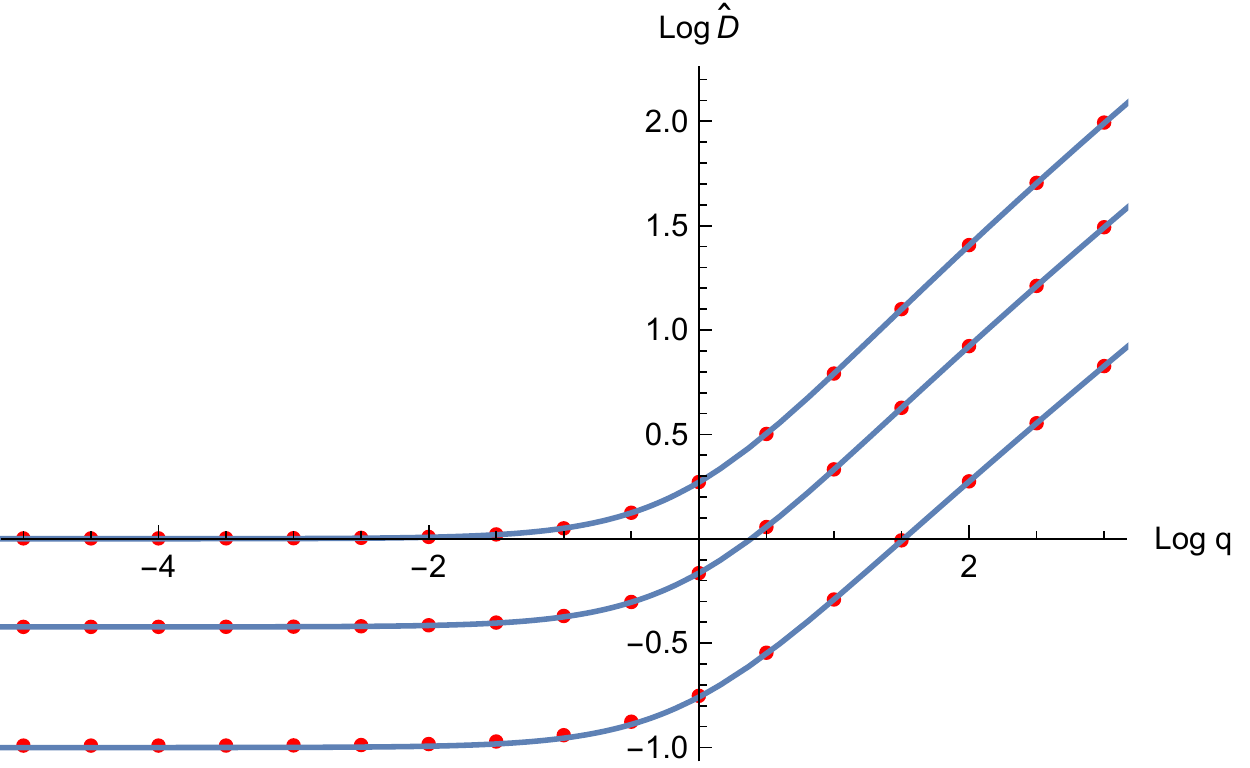}
 \caption{The numerical results (red points) for the diffusion constant $\hat D$ as a function of internal flux $q$ follow the analytic results (\ref{hat_diffusion}) (plotted in blue curves) spot on. The different curves correspond to varying magnetic field strength $\hat B=0,1,2$ (top-down) with $\hat d=10^6$.}
 \label{Diffusion_plot}
\end{figure}

Let us now consider the fluxed  D3-D5 system at non-zero temperature. First, we analyze the equations of the fluctuations near the horizon $r=r_h$. It is easy to verify that the equations decouple in this limit and that the equation for $E$ near $r=r_h$ takes the form:
\beq
E''\,+\,\Big({1\over r-r_h}\,+\,c_1\Big)\,E'\,+\,
\Big({A\over (r-r_h)^2}\,+\,{c_2\over r-r_h}\Big)\,E\,=\,0\,\,,
\label{E_eq_near-horizon}
\eeq
where $r_h$, $A$, $c_1$, and $c_2$ are constants.  Equation (\ref{E_eq_near-horizon}) can then be solved in a Frobenius series around $r=r_h$.  After this  near-horizon expansion we
perform a low frequency expansion by considering $k\sim {\mathcal O}(\epsilon)$, $\omega\sim {\cal O}(\epsilon^2)$. The coefficients $A$, $c_1$, and $c_2$ at leading order in $\epsilon$ are:
\bear
&& A\,=\,{\omega^2\over 16\,r_h^{2}} \rc
&& c_1\,=\,4(1+q^2)\,{r_h^3\over d^2+(1+q^2)(r_h^4+B^2)}\,{k^2\over \omega^2}\,+\,\ldots
\rc
&&c_2\,=\,-{1+q^2\over 4}\,{r_h\over d^2+(1+q^2)(r_h^4+B^2)}\,\,k^2
\,+\,\ldots\,\,.
\label{A_c1_c2}
\eear
It can be checked that near $r=r_h$, at leading order in $\epsilon$, the electric field $E(r)$ can be approximated as:
\beq
E\approx\,E_{nh}\,\big[1\,+\,\beta (r-r_h)\,\big]\,\,,
\label{nh_low_freq_E}
\eeq
where $E_{nh}$ is the value of $E$ at the horizon and  $\beta$ is a constant coefficient given by:
\beq
\beta\,=\,i\,{k^2\over \omega}\,(1+q^2)\,
{r_h^2\over d^2+(1+q^2)\,(r_h^4+B^2)}\,\,.
\eeq
We now perform the limits in opposite order. For low frequency, the equation of motion for $E$ can be written as:
\beq
E''\,-\,\partial_r\,\log\Bigg[{r^4+B^2\over
\sqrt{r^{4}+d^2+q^2\,r_h^4+(1+q^2)\,B^2}\,
\Big(d^2+(1+q^2)\,(r^{4}+B^2)\Big)}\Bigg]\,E'\,=\,0\,\,.
\eeq
This equation can be integrated as:
\beq
E(r)\,=\,E^{(0)}+ c_E\,{\cal I}(r)\,\,,
\label{diff_E_low_freq}
\eeq
where $E^{(0)}$ is the UV value of $E$,  $c_E$ is an integration constant, and  ${\cal I}(r)$ is the integral:
\beq
{\cal I}(r)\,=\,\int_{r}^{\infty}\,d\rho\,
{\rho^4+B^2\over 
\sqrt{\rho^{4}+d^2+q^2\,r_h^4+(1+q^2)\,B^2}\,
\Big(d^2+(1+q^2)\,(\rho^{4}+B^2)\Big)}\,\,.
\eeq
We now expand $E(r)$ in (\ref{diff_E_low_freq}) near the horizon:
\beq
E(r)\,=\,E^{(0)}+ c_E\,{\cal I}(r_h)\,-\,{(r_h^4+B^2)\,c_E\over 
\big[d^2+(1+q^2)\,(r_h^{4}+B^2)\big]^{{3\over 2}}}\,\,(r-r_h)\,+\,\ldots\,\,.
\label{low_freq_nh_E}
\eeq
Let us  now compare (\ref{nh_low_freq_E}) and (\ref{low_freq_nh_E}). From the comparison of the constant terms we arrive at:
\beq
E^{(0)}\,=\,E_{nh}-c_E\,{\cal I}(r_h)\,\,,
\eeq
while matching the linear terms yields:
\beq
c_E\,=\,-\,\beta\,{\big[d^2+(1+q^2)\,(r_h^{4}+B^2)\big]^{{3\over 2}}\over r_h^4+B^2}\,E_{nh}\,=\,
-i\,{k^2\over \omega}\,(1+q^2)
{r_h^2\over r_h^4+B^2}\,\big[d^2+(1+q^2)\,(r_h^{4}+B^2)\big]^{{1\over 2}}\,E_{nh}\,\,.
\eeq
Thus, we can write $E^{(0)}$ as:
\beq
E^{(0)}\,=\,E_{nh}\,\Big[1+
i\,{k^2\over \omega}\,{(1+q^2)\,r_h^{2}\over r_h^{4}\,+\,B^2}\,
\big[d^2+(r_h^{4}+B^2)\,(1+q^2)\big]^{{1\over 2}}
{\cal I}(r_h)\Big]\,\,.
\eeq
From the condition $E^{(0)}=0$  we get a dispersion relation of the type $\omega=-i D k^2$, with the diffusion constant $D$:
\beq
D\,=\,(1+q^2)\,{r_h^{2}\over  r_h^{4}\,+\,B^2}\,
\sqrt{d^2+(r_h^{4}+B^2)\,(1+q^2)}\,\,\,
{\cal I}(r_h)\,\,.
\label{Diffusion_D}
\eeq
In terms of the hatted variables defined in (\ref{rescaled_quantities}), the diffusive dispersion relation can be written as:
\beq
\hat \omega\,=\,-i\,\hat D\,\hat k^{2}\,\,,
\eeq
with the rescaled  diffusion constant $\hat D$ defined as $\hat D\,=\,r_h\,D\,=\,\pi\,T\,D$.  It is immediate   from (\ref{Diffusion_D}) to get the value of $\hat D$:
\beq
\hat D\,=\,(1+q^2)\,{\sqrt{\hat d^{\,2}+(1+\hat B^{\,2})(1+ q^{\,2})}\over
1+\hat B^{\,2}}\,\, {\cal J}(\hat d, \hat B, q)\,\,,
\label{hat_diffusion}
\eeq
where ${\cal J}(\hat d, \hat B, q)$ is the integral
\beq
{\cal J}(\hat d, \hat B, q)\,=\,\int_{1}^{\infty}\,dx\,
{x^{4}+\hat B^{\,2}\over
\sqrt{x^4+\hat d^{\,2}+ q^{\,2}+(1+ q^{\,2})\,\hat B^{\,2}}\,\,
\big[\hat d^{\,2}+(x^{4}+\hat B^{\,2})(1+ q^{\,2})\big]
}\,\,.
\eeq
In Fig.~\ref{Diffusion_plot}  we compare the numerical and analytical results for $\hat D$ as a function of $q$ for different values of the magnetic field $\hat B$.

Let us study some limits of the formulas for the diffusion constant we have just found. First of all, it is interesting to point out that the integral ${\cal J}$ can be performed analytically when $\hat d=0$. The resulting expression for $\hat D$ is:
\beq\label{eq:DD}
\hat D(\hat d=0)\,=\,\sqrt{{1+q^2\over 1+\hat B^2}}\,
F\Big({1\over 2},{1\over 4};{5\over 4};-q^2-(1+q^2)\,\hat B^2\Big)\,\,.
\eeq
As $\hat d\to 0$ when $T\to\infty$, the high temperature limit readily follows from (\ref{eq:DD}):
\beq
\lim_{T\to\infty}\,\hat D\,=\,\sqrt{1+q^2}\,\,F\big({1\over 2}, {1\over 4};{5\over 4};-q^2\big)\,\,.
\label{hatD_Tinfnity}
\eeq
Thus, at large $T$ the diffusion constant behaves as:
\beq
D\,\approx {\sqrt{1+q^2}\over \pi\,T}\,
\,F\big({1\over 2}, {1\over 4};{5\over 4};-q^2\big)\
\,\,,
\qquad\qquad
T\to\infty\,\,.
\label{D_large_T}
\eeq
Interestingly, this last expression coincides with the one found in \cite{Myers:2008me}. 
One can also study the opposite limit $T\to 0$. We find:
\beq
D\,\approx {\mu\over 2}\,{(1+q^2)\,(\pi T)^2\over (\pi T)^4+B^2}\,
\Bigg({d^2\over d^2+(1+q^2)\,B^2}\Bigg)^{{3\over 4}}\,
\Bigg[\,F\Big(1,{5\over 4};{3\over 2};-q^2\Big)\,+\,{2 B^2\over d^2}\Bigg]\,\,,
\qquad\qquad
T\to 0\,\,.
\qquad
\label{D_low_T}
\eeq

\section{Conclusions and outlook}
\label{conclusions}
In this paper we studied the collective excitations of cold holographic matter confined to a (2+1)-dimensional defect of $4d$ ${\cal N}=4$ super Yang-Mills theory in the Higgs branch. The string theory realization of the system is a D3-D5 intersection with flux on the worldvolume of the D5-brane.  We found a simple analytic expression for the dispersion relation of the zero sound as a function of the flux (see eqs. (\ref{Re_omega}) and (\ref{Im_omega})).  The speed of zero sound and the attenuation grow monotonically as the flux increases. We also studied the diffusion constant at higher temperatures. 

Our work can be naturally extended along several directions. We could study other observables of the fluxed D3-D5 system with general boundary conditions. Some of these observables are the AC and DC conductivities of the anyonic fluid, as well as its diffusion constant. Moreover, we could easily extend our results to the general D$p$-D$(p+2)$ intersections with flux.  

A more ambitious project could be the study collective excitations of the Higgs symmetry breaking even in more general holographic setups.  One of such systems could be the D3-D7 intersection with an instanton on the D7-brane worldvolume.  The explicit expression of this instanton at zero temperature and non-zero density has been found in \cite{Ammon:2012mu}.  Moreover, it was argued in \cite{Chen:2009kx} that this setup realizes holographically the color-flavor locking phase of color superconductivity.  The analysis of the collective excitations of this model is of obvious interest and we intend to address this problem in the near future.


\vspace{0.5cm}

{\bf \large Acknowledgments}

We thank Ra\'ul Arias, Yago Bea, and Javier Mas  for discussions. N.J. is supported by the Academy of Finland grant no. 1268023. 
G. I. and A.~V.~R. are  funded by the Spanish grant FPA2011-22594, by the Consolider-Ingenio 2010 Programme CPAN (CSD2007-00042), by Xunta de
Galicia (GRC2013-024), and by FEDER. We thank the Galileo Galilei Institute for Theoretical Physics for the hospitality and the INFN for partial support during the completion of this work.

\end{document}